\documentclass{moriond}
\usepackage[english]{babel}

\usepackage[backend=biber,sorting=none,style=numeric-comp,maxnames=1]{biblatex}%
\bibliography{references}
\AtEveryBibitem{\clearfield{title}}

\usepackage[symbol]{footmisc}

\usepackage{color,graphicx,epsfig}
\usepackage{amsmath,amssymb,amsfonts,dsfont,xcolor}
\usepackage{ifpdf}
\usepackage{bm}
\usepackage{braket}
\usepackage{hyperref}

\def\LjubljanaFMF{Faculty of Mathematics and Physics, University of Ljubljana,
Jadranska 19, 1000 Ljubljana, Slovenia }
\def\LjubljanaIJS{Jo\v zef Stefan Institute, Jamova 39, 1000 Ljubljana, Slovenia}
\def\Orsay{Universit\'e Paris-Saclay, CNRS/IN2P3, IJCLab, 91405 Orsay, France}
\newcommand{\Photo}{}

\begin{document}
\title{Implications of \boldmath$b\to s\ell^+\ell^-$ constraints on $b\to s\nu\bar\nu$ and $s\to d\nu\bar\nu$}

\author{S. Descotes-Genon$^{\, \mathrm{a}}$, S. Fajfer$^{\, \mathrm{b,c}}$,  J. F. Kamenik$^{\, \mathrm{b,c}}$,  {\bf M. Novoa-Brunet}$^{\, \mathrm{a}}$~\footnote[1]{Speaker}}

\address{
\vspace{4mm}
$^{\, \mathrm{a}}$\Orsay
\\[3pt]
$^{\, \mathrm{b}}$\LjubljanaIJS
\\[3pt]
$^{\, \mathrm{c}}$\LjubljanaFMF
}

\maketitle\abstracts{
We investigate the consequences of deviations from the Standard Model observed in $b\to s\mu\mu$ transitions for flavour-changing neutral-current processes involving down-type quarks and neutrinos, under generic assumptions concerning the structure of New Physics. We derive the relevant Wilson coefficients within an effective field theory approach respecting the SM gauge symmetry, including right-handed currents and assuming a flavour structure based on approximate $U(2)$ symmetry, and only SM-like light neutrinos. We discuss correlations among  $B \to K^{(*)} \nu \bar \nu$  and $K\to \pi \nu \bar \nu$ branching ratios
in the case of linear Minimal Flavour Violation and in a more general framework, highlighting in each case the role played by various New Physics scenarios proposed to explain $b\to s\mu\mu$ deviations. This talk is based on Ref.~\supercite{Descotes-Genon:2020buf}.}

\section{Introduction}\label{sec:introduction}
Recent experimental data in $B$ physics hint towards deviations from Lepton Flavour Universality (LFU) in semi-leptonic b-quark decays~\supercite{Bifani:2018zmi}.
These deviations can be interpreted model-independently in terms of specific contributions to short-distance Wilson coefficients $\mathcal{C}_i$ of the weak effective Hamiltonian \supercite{Buchalla:1995vs,Buras:1998raa}.
The global fits to $b\to s\ell\ell$ data \supercite{Alguero:2021anc} show that these deviations exhibit a consistent pattern favouring a significant additional New Physics (NP) contribution to $C^\mu_9$ (of the order of 25\% of the SM contribution) together with smaller contributions to $C^\mu_{10}$ and/or $C^\mu_{9'}$.
Among the most prominent scenarios~\supercite{Alguero:2021anc} one can find the one-dimensional scenarios $C^{\mu,{\rm NP}}_9= -1.06$, $C^{\mu,{\rm NP}}_9=-C^{\mu,{\rm NP}}_{10}=-0.44$ and $C^{\mu,{\rm NP}}_9=-C^{\mu,{\rm NP}}_{9'}=-1.11$.
Small contributions to electron operators $C^{e,{\rm NP}}_{9^{(\prime)},10^{(\prime)}}$ are allowed by the data, but they are not required to accurately describe the deviations. However, contributions to tau operators are mostly unconstrained \supercite{Capdevila:2017iqn}.

The SM neutrinos reside in the same leptonic weak doublets as the left-handed charged leptons. Therefore, decay modes with neutrinos in the final state offer complementary probes of NP.  In particular, decays $B\to h_s \nu \bar \nu$, with $h_s$  standing  for  hadronic states of strangeness equal to 1, are known for their NP sensitivity~\supercite{Altmannshofer:2009ma}.
Another related and particularly interesting question is whether NP is  only present in $b \to s$ transitions or also in other Flavour-Changing Neutral Currents (FCNC).  The  measurements of $s\to d \nu \bar \nu$  and $b\to s \nu \bar \nu$  rates should help to differentiate  among NP  models with different flavour and chiral structures in both the quark and lepton sectors.
In the kaon sector $K \to \pi \nu \bar \nu$  decays arguably offer the best sensitivity to NP~\supercite{Buras:1998raa}.
The main goal of our  approach is to determine the impact of the current $b\to s\ell\ell$ results on future measurements of $B \to K^{(*)} \nu \bar \nu$ and $K\to \pi \nu \bar \nu$ in a general effective theory framework and to illustrate the potential correlations among these measurements.

\section{NP in semileptonic FCNC decays}
Possible heavy NP contributions are written in terms of $SU(2)_L$ gauge invariant operators~\supercite{Buttazzo:2017ixm}
\begin{equation}
  \begin{split}
    & \mathcal L_{\rm eff.}   = \mathcal L_{\rm SM}  - \frac{1}{v^2} \lambda^q_{ij} \lambda^\ell_{\alpha\beta} \left[ C_T \left( \bar Q_L^i \gamma_\mu \sigma^a Q_L^i \right) \left( \bar L_L^\alpha \gamma^\mu \sigma^a L_L^\beta \right) +  C_S \left( \bar Q_L^i \gamma_\mu  Q_L^i \right) \left( \bar L_L^\alpha \gamma^\mu  L_L^\beta \right)   \right.\\
    &\left.  + C'_{RL} \left( \bar d_R^i \gamma_\mu d_R^i \right) \left( \bar L_L^\alpha \gamma^\mu  L_L^\beta \right) + C'_{LR} \left( \bar Q_L^i \gamma_\mu  Q_L^i \right) \left( \bar \ell_R^\alpha \gamma^\mu  \ell_R^\beta \right)  + C'_{RR} \left( \bar d_R^i \gamma_\mu d_R^i \right) \left( \bar \ell_R^\alpha \gamma^\mu  \ell_R^\beta \right)
    \right]\,, \label{eq:ops}
  \end{split}
\end{equation}
where $Q_L^i = (V^{\rm CKM *}_{ji}  u_L^j, d_L^i)^T$
and $L_L^\alpha = (U^{\rm PMNS}_{\alpha\beta}\nu_L^\beta,\ell^{\alpha}_L)^T$.
This Lagrangian, extending the one considered in Ref.~\supercite{Buttazzo:2017ixm} to include right-handed fields, is chosen to reproduce the results of the global fits to $b\to s\ell\ell$ data\footnote{$b\to s\ell\ell$ data suggests scenarios limited to axial and vector NP ($\mathcal O^\ell_{9(^\prime),10(^\prime)}$), without scalar or tensor NP.} discussed in Sec.~\ref{sec:introduction}.
The presence of operators with lepton doublets in Eq.~(\ref{eq:ops}) is the basis for the connection between flavour-changing neutral currents involving charged and neutral leptons that we explore in the following.
We assume that the same flavour structure encoded in $\lambda^q_{ij}$ and $\lambda^\ell_{\alpha\beta}$ holds for all operators.
We then classify the NP flavour structures in terms of an approximate  $U(2)_{q=Q,D}$ flavour symmetry acting directly on the quark fields, under which two generations of quarks form doublets, while the third generation is invariant. We will focus on down-type quarks. One can write
${\bf q}\equiv (q_L^1, q_L^2) \sim ({\bf 2},{\bf 1})$, ${\bf d} \equiv (d_R^1,d_R^2) \sim ({\bf 1},{\bf 2})$ while $ d_R^3,q_L^3 \sim ({\bf 1},{\bf 1})$\,.
In the exact $U(2)_{q}$ limit only $\lambda^q_{33}$ and $\lambda^{q}_{11}=\lambda^q_{22} $ in Eq.~\eqref{eq:ops} are non-vanishing.
To avoid excessive effects in neutral kaon oscillation observables, we thus furthermore impose the {\it leading} NP $U(2)_q$ breaking to be aligned with the SM Yukawas, yielding a General Minimal Flavour Violating (GMFV)~\supercite{Kagan:2009bn} structure with the singlet field defined as
$d _{L}^3 =  b_L+ \theta_q e^{i\phi_q} \left( V_{td} d_L   + V_{ts}  s_L \right)$, whereas $d^1_L=d_L$, $d^2_L=s_L$ (and similarly for $q^3_L$), where $\theta_q$ and $\phi_q$ are fixed but otherwise arbitrary numbers.
The linear MFV limit~\supercite{Kagan:2009bn} is recovered by taking $\theta_q = 1$ and $\phi_q=0$ (taking $V_{tb}=1$)~\footnote{Note that, because of the rigid flavour breaking structure within the linear MFV regime, the flavour construction in Eq.~\eqref{eq:ops} does not imply extra assumptions, any additional flavour re-scalings of individual operators can be absorbed into the flavour universal $C_i$'s.}.
In (G)MFV,  right-handed FCNCs among down-type quarks are suppressed so that we may set $C'_{RL}=C'_{RR}=0$ then.
Departures from the (G)MFV limit may manifest through additional
explicit $U(2)_q$ breaking effects appearing as $\lambda^q_{i\neq j} \neq 0$.

For the lepton sector we assume an approximate $U(1)^3_\ell$ symmetry (broken only by the neutrino masses) yielding $\lambda_{i\neq j}^\ell \simeq 0$\, as required by stringent limits on lepton flavour violation~\supercite{Davidson:2006bd,Glashow:2014iga,Alonso:2015sja}.
We consider here only (SM-like) left-handed neutrinos. Since neutrino flavours are not tagged in current and upcoming rare meson decay experiments, we need to assume specific ratios of $U(1)_\ell^3$ charges ($\lambda^{\ell}$) in order to correlate FCNC processes involving charged leptons and neutrinos. We consider three well known examples from the existing literature.
First, the simplest (scenario 1) $\lambda^{\ell}_{ee} =\lambda^{\ell}_{\tau\tau} = 0$ implies significant NP effects only in muonic final states.
Secondly, the anomaly-free assignment (scenario 2) $\lambda^{\ell}_{\mu\mu} = -\lambda^{\ell}_{\tau\tau}$ and $\lambda^{\ell}_{ee} =0$ allows for gauging of the leptonic flavour symmetry and is thus well suited for UV-complete model building~\supercite{Altmannshofer:2014cfa,Crivellin:2016ejn}.
And lastly, the hierarchical charge scenario (scenario 3) $\lambda^{\ell}_{ee} \ll \lambda^{\ell}_{\mu\mu} \ll \lambda^{\ell}_{\tau\tau}$
\footnote{For concreteness we consider $\lambda^{\ell}_{\alpha\alpha} / \lambda^{\ell}_{\mu\mu} = m_\alpha / m_\mu$ with $\alpha=e,\,\tau$.} which is motivated by models of partial lepton compositeness and flavour models accounting for hierarchical charged lepton masses~\supercite{Redi:2011zi,Niehoff:2015bfa}.
The expressions of the branching fractions and the weak effective Hamiltonian Wilson coefficients as a function of the coefficients of Eq.~\eqref{eq:ops} can be found in Ref.~\supercite{Descotes-Genon:2020buf}.

\section{Results} 
\begin{figure}[h]
    \centering
    \includegraphics[width=0.45\textwidth]{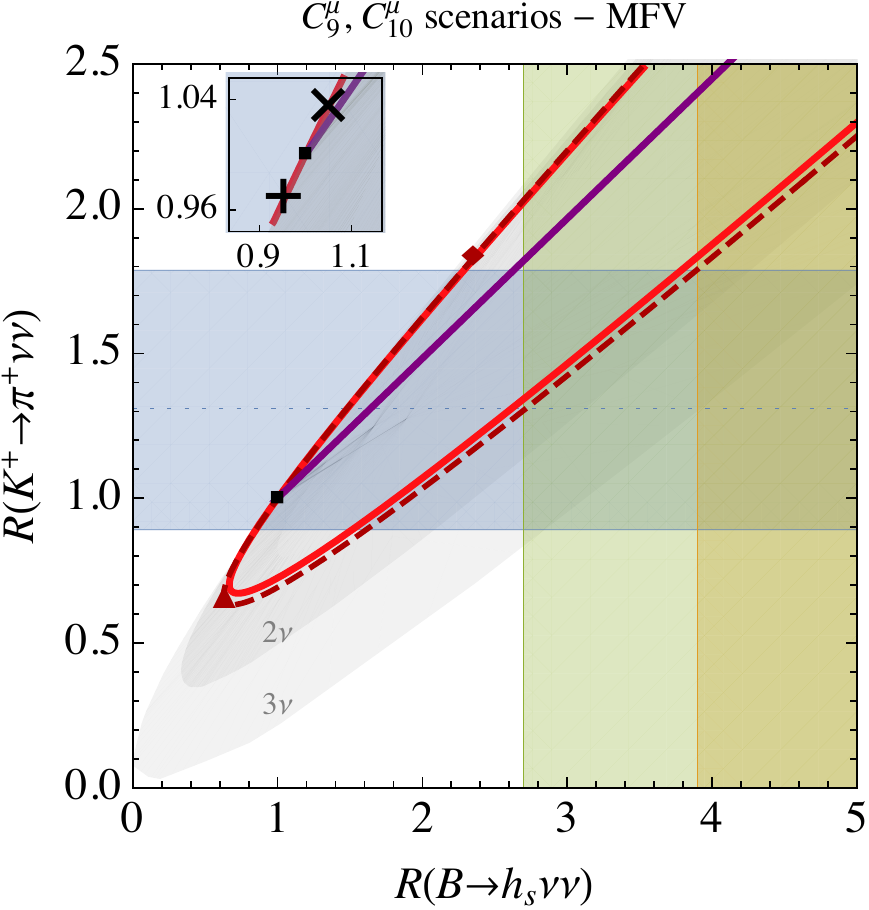}
    \hspace{3em}
      \includegraphics[width=0.45\textwidth]{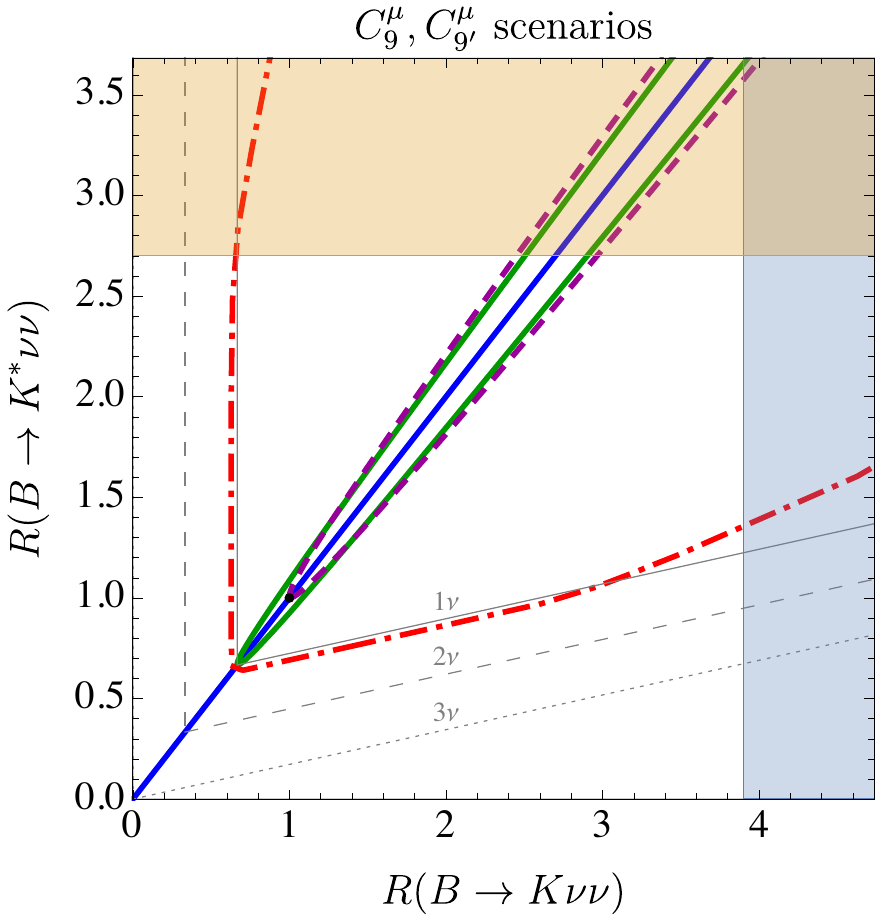}
    \caption{Left: correlation between the ratios $R(B\to h_s\nu\bar\nu)$ ($h_s=K,K^*,X_s$) and $R(K^+\to \pi^+\nu\bar\nu)$ in the MFV limit.
    Right: correlation between the ratios  $R(B\to K\nu\bar\nu)$ and $R(B\to K^*\nu\bar\nu)$ in presence of right-handed NP in  $b\to s\mu\mu$ $(C^{\mu,\rm NP}_9, C^{\mu,\rm NP}_{9'})$.
    Regions are explained in the text.}
    \label{fig:C9C10}
  \end{figure}

We first consider the limit of (linear) MFV in which $b\to s\nu\bar\nu$ and $s\to d\nu\bar\nu$ FCNC transitions are rigidly correlated through their dependence on $C_S-C_T$, even before considering the implications of $b\to s\ell^+\ell^-$.
In Fig.~\ref{fig:C9C10} (left) we show the branching ratio of $B\to h_s \nu\nu$ and $K^+\to \pi^+\nu\bar\nu$ normalised to their respectively SM value  $R(i\to f) \equiv \mathcal B(i\to f) /  \mathcal B(i\to f)_{\rm SM}$.
The allowed region for these ratios is shown shaded in Fig.~\ref{fig:C9C10} (left) for arbitrary MFV NP effects on two (dark grey, 2$\nu$) or three (light grey, 3$\nu$) neutrino flavours. In the case of the three specific $U(1)^3_\ell$ scenarios, the red, purple and dashed brown curves correspond to the allowed 1D regions for Sc. 1, 2 and 3 respectively.
In the $b \to s \ell^+ \ell^-$ analysis, the MFV limit corresponds to the $(C_9^{\mu,\rm NP}$, $C_{10}^{\mu,\rm NP})$ scenario or in terms of the operators in Eq.~\eqref{eq:ops} $(C_S+C_T,C'_{LR})$.
Since $B \to h_s \nu\bar\nu$ and $K\to\pi \nu \bar\nu$ depend on the orthogonal $C_S-C_T$ combination, interesting implications can only be derived in specific scenarios allowing us to convert the information from $b \to s \ell^+ \ell^-$ observables into a constraint on $C_S$ and $C_T$.
The simplest possibilities $C_S=0$ or $C_T=0$ are indicated as black $\times$ and $+$ respectively in the inset plot for scenario 1 and brown $\Diamond$ or $\bigtriangleup$ respectively  for scenario 3 (already being probed by $\mathcal B(K^+\to \pi^+\nu\bar\nu)$ measurement and close to being probed by $B \to K^{(*)} \nu\bar\nu$ at the $B$-factories). On the other hand, in scenario 2, no significant deviations are expected in either case.
Both ratios $R$ are bounded by the same minimal value $(1-N_\nu/3)$ where $N_\nu$ is the number of neutrino flavours affected by NP.
Also shown are the present experimental constraints coming from NA62~\supercite{CortinaGil:2021nts} and $B$-factories~\supercite{Grygier:2017tzo}.
An interesting observation is that a pair of future $B \to h_s \nu\bar\nu$ and $K\to\pi \nu \bar\nu$ rate measurements outside of this (albeit large) region would be a clear indication of non-MFV NP.

  Beyond the linear MFV limit any correlation between $b\to s$ and $s\to d$ FCNCs is lost. Nonetheless, the potential presence of right-handed $b\to s$ FCNCs in the $(C^{\mu,\rm NP}_9, C^{\mu,\rm NP}_{9'})$ scenario\footnote{Note that the additional inclusion of $C^{\mu,\rm NP}_{10}$ to this NP scenario does
  not alter our conclusions, as it corresponds to the addition of $C'_{LR}$ which is not involved in $b\to s\nu\bar\nu$.} as well as the leptonic flavour structure of NP can both still be probed using correlations among two $B \to h_s \nu\bar\nu$  modes, as shown in Fig.~\ref{fig:C9C10} (right) for the case $R(B \to K^{} \nu\bar\nu)$ vs. $R(B \to K^{*} \nu\bar\nu)$.
  The diagonal blue line corresponds to the (G)MFV case ($C^{\mu,\rm NP}_{9'}=0$).
  In scenario 1 and scenario 2 the $b\to s\ell\ell$ fit 1$\sigma$ region for $(C^{\mu,\rm NP}_9, C^{\mu,\rm NP}_{9'})$ singles out a narrow region (inside the solid green and dashed purple lines respectively) around the diagonal in this plane, whereas scenario 3 (inside dot-dashed red line) leaves a much larger region allowed.
  Conversely, a measurement of the two $b\to s \nu\bar \nu$ modes outside of the region for scenario 1 would indicate significant (right-handed FCNC) NP couplings to other neutrino species, e.g. $\nu_\tau$. Without information on the size of the right-handed FCNCs from $b\to s \mu^+ \mu^-$, the allowed region assuming significant NP couplings to 1, 2, 3 neutrinos is above and on the right of the solid, dashed, dotted grey contours, respectively. The horizontal and vertical bands correspond to the 90 \% CL limits on the observables for $R(B\to K^*\nu\bar\nu)$ (orange) and $R(B\to K\nu\bar\nu)$ (blue).
  First note that in the MFV limit relative NP effects in both modes are expected to be identical as indicated by the  diagonal blue line. Beyond MFV however, any deviation from the diagonal would indicate the presence of right-handed currents and the amount of deviation from the diagonal would directly indicate the number of lepton flavours affected by NP.

  In summary, we have investigated, in a general EFT framework, the consequences of  $b\to s\mu\mu$ SM deviations in other FCNC processes. Under specific assumptions, we have studied the correlation between the branching ratios for $B\to h_s\nu\bar\nu$ and $K^+\to\pi^+\nu \bar\nu$ and between $B\to K\nu\bar\nu$ and $B\to K^*\nu\bar\nu$. The measurement of this modes could establish NP flavour breaking beyond (G)MFV as well as indicating the number of lepton flavours affected by NP.

  \section*{Acknowledgements}
  SF and JFK acknowledge the financial support from the Slovenian Research Agency (research core funding No. P1-0035 and J1-8137). This project has received support from the European Union’s Horizon 2020 programme (grant agreement No 860881-HIDDeN).

  \printbibliography

  \end{document}